\documentclass[10pt,journal,compsoc]{IEEEtran}
\ifCLASSOPTIONcompsoc
  \usepackage[nocompress]{cite}
\else
  \usepackage{cite}
\fi
\ifCLASSINFOpdf
\else
\fi

\usepackage{booktabs} 
\usepackage{subfig}
\usepackage{graphicx}
\usepackage{algorithm}
\usepackage{algpseudocode}%
\usepackage{xcolor}
\usepackage{amsmath, amsthm}
\usepackage{amssymb}
\newcommand{\etal}{\textit{et al}. }

\usepackage{mathtools}
\usepackage{multirow}
\usepackage{graphics}
\usepackage{enumitem}
\usepackage{CJKutf8}
\usepackage{array}
\usepackage{diagbox}
\usepackage{svg}
\usepackage{threeparttable}
\usepackage{comment}
\usepackage{xcolor,cite,etoolbox}
\definecolor{mypurple}{rgb}{0.4392, 0.1882, 0.6275}
\definecolor{darkblue}{rgb}{0.0, 0.0, 0.55}
\definecolor{darkgray}{rgb}{0.66, 0.66, 0.66}
\makeatletter 
\pretocmd\@bibitem{\color{black}\csname keycolor#1\endcsname}{}{\fail}
\newcommand \citecolor[1]{\@namedef{keycolor#1}{\color{red}}}
\makeatother

\begin{document}
%
\title{DuA: Dual Attentive Transformer in\\ Long-Term Continuous EEG Emotion Analysis}
%
%
%
%

\author{\IEEEauthorblockN{Yue Pan\textsuperscript{1}, Qile Liu\textsuperscript{1}, Qing Liu, Li Zhang, Gan Huang,\\ Xin Chen, Fali Li\textsuperscript{*}, Peng Xu\textsuperscript{*}, and Zhen Liang\textsuperscript{*}, \IEEEmembership{Member, IEEE}}\\
\medskip
\IEEEcompsocitemizethanks{\IEEEcompsocthanksitem Yue Pan is with the School of Biomedical Engineering, Medical School, Shenzhen University, Shenzhen 518060, China, with the Guangdong Provincial Key Laboratory of Biomedical Measurements and Ultrasound Imaging, Shenzhen 518060, China, with The Clinical Hospital of Chengdu Brain Science Institute, MOE Key Lab for Neuroinformation, University of Electronic Science and Technology of China, China, and also with School of Life Science and Technology, Center for Information in Medicine, University of Electronic Science and Technology of China, China. (e-mail: yue.pan@std.uestc.edu.cn).
\IEEEcompsocthanksitem Qile Liu, Li Zhang, Gan Huang, Xin Chen, and Zhen Liang are with the School of Biomedical Engineering, Medical School, Shenzhen University, Shenzhen 518060, China, also with the Guangdong Provincial Key Laboratory of Biomedical Measurements and Ultrasound Imaging, Shenzhen 518060, China. (e-mail: liuqile2022@email.szu.edu.cn, lzhang@szu.edu.cn, huanggan@szu.edu.cn, and janezliang@szu.edu.cn).
\IEEEcompsocthanksitem Qing Liu is with the Rehabilitation Medicine Department, Shenzhen Children's Hospital, China (e-mail: liuqing-golden@163.com).
\IEEEcompsocthanksitem Fali Li, and Peng Xu are with The Clinical Hospital of Chengdu Brain Science Institute, MOE Key Lab for Neuroinformation, University of Electronic Science and Technology of China, China, and with School of Life Science and Technology, Center for Information in Medicine, University of Electronic Science and Technology of China, China. (e-mail: fali.li@uestc.edu.cn, xupeng@uestc.edu.cn).
}}
\IEEEtitleabstractindextext{%
\begin{abstract}
Affective brain-computer interfaces (aBCIs) are increasingly recognized for their potential in monitoring and interpreting emotional states through electroencephalography (EEG) signals. Current EEG-based emotion recognition methods perform well with short segments of EEG data (known as segment-based emotion analysis). However, these methods encounter significant challenges in real-life scenarios where emotional states evolve over extended periods. To address this issue, we propose a \textbf{Du}al \textbf{A}ttentive (DuA) transformer framework for long-term continuous EEG emotion analysis. Unlike segment-based approaches, the DuA transformer processes an entire EEG trial as a whole, identifying emotions at the trial level, referred to as trial-based emotion analysis. This framework is designed to adapt to varying signal lengths, providing a substantial advantage over traditional methods. The DuA transformer incorporates three key modules: the spatial-spectral network module, the temporal network module, and the transfer learning module. The spatial-spectral network module simultaneously captures spatial and spectral information from EEG signals, while the temporal network module detects temporal dependencies within long-term EEG data. The transfer learning module enhances the model's adaptability across different subjects and conditions. We extensively evaluate the DuA transformer using a self-constructed long-term EEG emotion database, along with two benchmark EEG emotion databases. On the basis of the trial-based leave-one-subject-out cross-subject cross-validation protocol, our experimental results demonstrate that the proposed DuA transformer significantly outperforms existing methods in long-term continuous EEG emotion analysis, with an average enhancement of 5.28\%. The DuA transformer's ability to adapt to varying signal lengths and its superior performance across diverse subjects and conditions highlight its potential for real-world applications, enhancing the overall user experience and efficacy of aBCI systems.
\end{abstract}

}

\maketitle

\footnotetext[1]{\hspace{1mm}Equal contributions. \textsuperscript{*} Corresponding authors.}
\IEEEdisplaynontitleabstractindextext

%
\IEEEpeerreviewmaketitle

\IEEEraisesectionheading{
\section{Introduction}
\label{sec:introduction}}
\IEEEPARstart{E}{lectroencephalography} (EEG), a non-invasive technique for monitoring brain electrophysiological activity, records neuronal electrical activity signals through scalp electrodes. Compared to other physiological signals, EEG more directly reflects changes in emotions, providing a neuroscientific interpretation of emotional states \cite{picardAffectiveComputing2000,liang2019unsupervised}. For certain emotional states, a long-term continuous evolution may be necessary. Take the sadness as an example. Short stimuli might not effectively evoke vivid sadness, nor might they induce intense, profound sadness in subjects or bring about significant changes in brain activity. This limitation can lead to poor results in emotion recognition. Saarimäki \etal \cite{saarimakiClassificationEmotionCategories2022} found the lowest average classification accuracy for sadness (18\%) in their study. In Raz \etal's work\cite{razPortrayingEmotionsTheir2012}, it demonstrated that long and complex movie clips are more suitable for inducing the dynamic changes associated with sustained emotional experiences. Furthermore, Xu \etal \cite{xuFunctionalConnectivityProfiles2023} suggested that a 10-minute long stimulus could be more beneficial for brain signal analysis, closely mirroring sustained emotional experiences. This finding confirms that emotional representations induced by long-duration stimuli are superior to those induced by short-duration stimuli. 

However, most existing EEG-based emotion studies utilize short-duration stimuli to evoke responses \cite{zhengInvestigatingCriticalFrequency2015, koelstraDEAPDatabaseEmotion2012}. The algorithms used in these studies typically perform emotion analysis on segmented EEG signals (\textbf{segment-based emotion analysis}), rather than processing an entire trial as a whole (\textbf{trial-based emotion analysis}) \cite{zhou2023pr, zhongEEGBasedEmotionRecognition2022,taoEEGBasedEmotionRecognition2023}. A more detailed introduction to these current algorithms is provided in Section \ref{sec:relatedWork}. This approach neglects the temporal information of the entire signal sequence, thus impacting the recognition performance on the whole EEG signal. Implementing a long-duration, whole-segment EEG emotion recognition algorithm faces the challenge of handling variable-length data inputs.

The transformer network excels in processing variable-length data and has been increasingly applied in the field of EEG for emotion analysis \cite{taoEEGBasedEmotionRecognition2023, jiaSSTEmotionNetSpatialSpectralTemporalBased2020, sunDualBranchDynamicGraph2022}. For example, Liu \etal \cite{liu2022spatial} proposed four variant transformer frameworks (spatial attention, temporal attention, sequential spatial-temporal attention and simultaneous spatial-temporal attention) for EEG-based emotion recognition, exploring the relationship between emotion properties and EEG features. Similarly, Sun \etal \cite{sunDualBranchDynamicGraph2022} introduced a transformer-based dynamic graph convolutional neural network (CNN) designed for feature fusion, where the extracted graph features are further updated by the transformer. Additionally, Wei \etal \cite{wei2023tc} proposed a transformer capsule network that integrates an EEG transformer module to extract EEG features and an emotion capsule module to refine these features for easier classification. However, existing transformer-based EEG algorithms have several notable limitations, such as inadequate incorporation of spatial and spectral information, insufficient handling of long-term dependencies, and a lack of research on long-term EEG modeling methods. To tackle the existing limitations, we propose a novel DuA transformer framework for long-term continuous EEG emotion analysis. The main contributions of this study are summarized as below.

\begin{itemize}
    \item We propose a novel \textbf{Du}al \textbf{A}ttentive (DuA) transformer framework for long-term continuous EEG analysis. This novel framework enhances traditional attention mechanisms by improving the performance of emotion decoding for long sequential signals, reducing GPU memory usage, and effectively utilizing the spatial-spectral and temporal information of EEG signals.
    \item We approach emotion analysis with a trial-based interpretation. Unlike traditional methods that segment EEG trials into short, fixed-length segments (e.g., 1 second) and assign the same label to all segments within a single trial, our proposed framework processes entire EEG trials with variable lengths. This allows for a more flexible and comprehensive analysis of long-term continuous EEG data.
    \item Extensive experiments are conducted on self-constructed long-term EEG database and two benchmark databases, covering various data lengths and emotional states. The model's reliability and effectiveness are validated using a strict trial-based leave-one-subject-out cross-subject cross-validation protocol, resulting in an average performance enhancement of 5.28\%.
\end{itemize}

\begin{figure*}
    \begin{center}
    \includegraphics[width=0.9\textwidth]{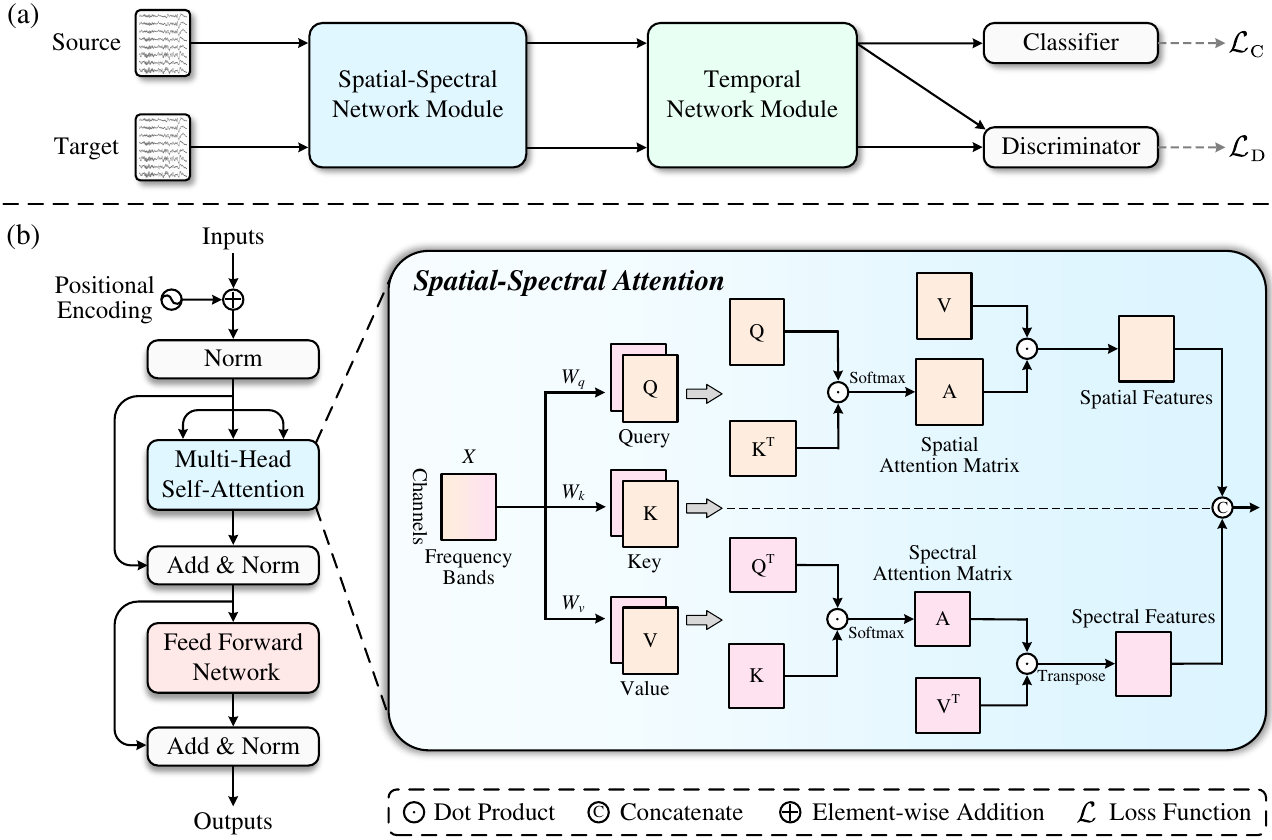}
    \end{center}
    \caption{The overall architecture of the proposed Dual Attentive (DuA) transformer model. (a) the main pipeline of the model, and (b) details the structure of the spatial-spectral network module.}
\label{Fig:Framework of Long-Term Transformer Network}
\end{figure*}

\section{Related Work}
\label{sec:relatedWork}

\subsection{Short Segments based EEG Analysis}
In recent years, a growing number of deep learning algorithms have been developed for EEG-based emotion recognition. For example, Liang \etal \cite{liangEEGFuseNetHybridUnsupervised2021} introduced EEGFuseNet to dynamically capture both shallow and deep EEG features without requiring label information. Fernandez \etal \cite{fdezCrossSubjectEEGBasedEmotion2021} leveraged prior knowledge to train a CNN using differential entropy (DE) features, improving the network's ability to recognize emotions. To incorporate structural information, Zhong \etal \cite{zhongEEGBasedEmotionRecognition2022} developed a restricted graph network with NodeDAT and EmotionDL, achieving an impressive accuracy of 85.30\% in 3-class cross-subject emotion recognition. However, most existing EEG-based emotion recognition algorithms rely on segmenting EEG trials into fixed-length segments (e.g., 1 second), with each segment assigned the same emotional label \cite{koelstraDEAPDatabaseEmotion2012, zhengIdentifyingStablePatterns2019, zhengEmotionMeterMultimodalFramework2019}. This approach assumes that the emotional state remains constant throughout each segment in a single trial. 

In practical emotion experiments, labels are typically provided for entire trials rather than individual segments. Emotional states are dynamic and can fluctuate significantly over longer periods, making it unrealistic to assume a static emotional label for each short segment. Moreover, research has shown that emotion elicitation is an accumulative process, evolving rather than remaining fixed \cite{wangClassificationUnmedicatedBipolar2020}. Given these considerations, emotion analysis using EEG signals should be conducted in a trial-based manner rather than a segment-based manner, especially in supervised learning contexts where accurate label information is crucial. This trial-based approach acknowledges the dynamic nature of emotional states and allows for more accurate and realistic emotion recognition in continuous EEG analysis.

\subsection{Transfer Learning in aBCI}
Emotion recognition algorithms face significant challenges due to the variability and complexity of EEG signals and the individual differences in these signals. To address these issues, transfer learning has emerged as a valuable technique, helping to mitigate the discrepancies in feature distribution extracted from various subjects and enhancing model stability in cross-subject EEG-based emotion recognition tasks. 

For non-deep transfer learning methods, Zheng \etal \cite{zheng2016personalizing} employed transfer component analysis (TCA) and transductive parameter transfer (TPT) to achieve an average accuracy of 76.31\%, compared to 56.73\% without transfer learning. Considering the marginal and conditional distributions in the feature alignment process, Luo \textit{et al.}\cite{luo2024mddd} introduced a manifold-based domain adaptation method. This method adaptively and dynamically adjusted the importance of marginal and conditional distributions during the feature alignment process based on Grassmann manifold space, leading to a further average improvement of 3.54\%.

Deep transfer learning algorithms have also shown promise in this field. Jin \etal \cite{jin2017eeg} introduced an EEG-based emotion recognition model using domain adversarial neural networks (DANN) The results demonstrated that models using deep transfer learning frameworks outperformed those using non-deep transfer methods, with average accuracy increasing from 76.31\% to 79.19\%. To further enhance feature alignment, a series of improved DANN frameworks were proposed \cite{he2022adversarial, peng2022joint}. Considering the interaction feature representation between sample features and prototype features, Zhou \etal \cite{zhou2023pr} introduced a prototype-representation pairwise learning-based transfer learning framework. This approach not only addressed individual differences but also tackled the issue of noisy labeling, achieving state-of-the-art (SOTA) results in cross-subject emotion recognition. The above studies demonstrate that transfer learning, both non-deep learning and deep learning, plays a crucial role in advancing aBCI systems by improving the generalization and robustness of emotion recognition models across different individuals. Future research should continue to explore optimized transfer learning methods, aiming to further bridge the gap between subject-specific models and generalized emotion recognition frameworks.

\subsection{Transformer Network with Attention Mechanism}
Transformer networks, introduced by Vaswani \etal \cite{vaswaniAttentionAllYou2017}, have become integral in both computer vision and natural language processing. In computer vision, models such as vision transformers (ViT) and swin transformers have set new standards. In natural language processing, models like BERT and GPT exemplify their success. The strength of transformer networks lies in their extensive use of attention mechanisms to process sequential data. This mechanism enables the model to identify and focus on the most relevant parts of the input. For each input element, the model assigns a weight indicating its importance, and through learning, it adaptively adjusts these weights. This process mimics human attention to specific details, significantly enhancing the model's performance.

In the field of continuous EEG analysis, a number of studies have integrated attention mechanisms and transformer networks to enhance performance. For example, Jia \etal \cite{jiaSSTEmotionNetSpatialSpectralTemporalBased2020} proposed a 3D fully-connected network based on attention mechanisms for DE feature projection. Tao \etal \cite{taoEEGBasedEmotionRecognition2023} incorporated self-attention mechanisms into CNNs and long short-term memory (LSTM) networks for better feature representation. Additionally, Si \etal \cite{si2023temporal} introduced a temporal aware mixed attention-based convolution and transformer network, which combines CNN and attention mechanisms to jointly extract local and global emotional features of EEG segments. Furthermore, some researchers attempts to directly apply transformer networks. Wang \etal \cite{wang2022temporal} proposed a transformer-based model to robustly capture temporal dynamics and spatial correlations of EEG signals. Sun \etal \cite{sunDualBranchDynamicGraph2022} extracted power spectral density (PSD) and DE features and then utilized graph and transformer networks to fuse and extract features representing emotions. Similarly, Cui \etal \cite{cui2024mvgt} introduced a multi-view graph transformer based on spatial relations that integrates information from the temporal frequency and spatial domains to enhance the expressive power of the model comprehensively. However, most existing methods emphasise learning short-term temporal patterns, neglecting significant long-term contextual and information related to emotional cognitive processes \cite{ding2024emt}. In addition, these methods inadequate incorporate spatial and spectral information, further leading to poor model performance in long-term continuous EEG analysis tasks.

To address the challenge of processing long-term continuous EEG signals for emotion recognition, we propose a Dual Attention (DuA) transformer framework, as illustrated in Fig. \ref{Fig:Framework of Long-Term Transformer Network}. Unlike typical one-dimensional time series, long-term EEG signals encompass rich temporal and spatial information. Directly applying transformer networks to these time series might overlook the spatial nuances, thereby diminishing the network's ability to accurately represent features and impacting the effectiveness of emotion recognition. On the other hand, traditional transformer networks, consisting of encoders and decoders, are typically employed for diverse tasks. Encoders are generally used for classification, while decoders are suited for generation tasks. Given that EEG-based emotion recognition is fundamentally a classification task, our approach focuses on utilizing the encoder architecture of the transformer network. 

Our proposed framework comprises three main modules: the spatial-spectral network, the temporal network, and a transfer learning module. This model extracts spatiotemporal features from EEG signals, fully exploiting both the temporal and spatial dimensions, overcoming the limitations inherent in traditional transformer networks. Additionally, to address the variability in EEG signals across different individuals, we incorporate a transfer learning strategy, enhancing the network's accuracy in cross-subject long-term EEG emotion recognition tasks. This combination of modules ensures a comprehensive analysis of trial-based EEG signals, leading to more accurate and reliable continuous EEG emotion analysis.

\section{Methodology}
\label{sec:model}
To address the limitations of traditional transformer networks in effectively utilizing the diverse aspects of long-term EEG signals, we propose a DuA transformer network specifically designed for continuous EEG analysis. DuA transformer includes three key modules. \textbf{Spatial-Spectral Network Module.} This module targets the spatial domain by analyzing the spatial distribution of EEG signals. It captures intricate spatial relationships between different EEG channels in terms of spectral patterns, enabling the model to understand complex spatial dependencies. \textbf{Temporal Network Module.} This module focuses on the temporal domain, effectively modeling the dynamic changes in EEG signals over time. By doing so, it captures the temporal dependencies and variations inherent in EEG data, allowing for a more nuanced understanding of the temporal dynamics. \textbf{Transfer Learning Module.} This module adapts the model to new individuals by leveraging knowledge from previously seen data, ensuring better generalization across different individuals. It enhances the model's ability to generalize and perform well on unseen subjects by transferring learned features and patterns from prior data. The details of each module will be illustrated below.

\subsection{Transformer Basics}
In this section, we will introduce the encoder part of the transformer network and its various configurations. The input sequence is processed through the encoder to extract useful information and project the sequence into low-dimensional features for subsequent downstream tasks, such as classification. The encoder consists of several stacked encoder layers. The input and output of the first layer are sequences, and each subsequent layer's input is the output from the previous layer. Each encoder layer is composed of a multi-head self-attention (MHSA) mechanism and a feed-forward network (FFN), connected by residual connections for deeper network construction, followed by layer normalization (LN).

The MHSA mechanism consists of several self-attention units. Self-attention operates on a "matching" mechanism, mapping each element of the sequence into Key, Query, and Value vectors. It calculates the dot product between the Query and Key vectors, transforming the result into attention scores ranging from 0 to 1 using the Softmax function. These attention scores are then multiplied with the Value vectors as follows:
\begin{equation}
    \label{eq:self attention}
        \begin{array}{l}
            \left\{\begin{matrix}   
            q = W_{q} \times X,q\in R^{n\times d_{k}},\\  
            k = W_{k} \times X,k\in R^{n\times d_{k}},\\  
            v = W_{v} \times X,v\in R^{n\times d_{k}},
            \end{matrix}\right.\\
        \end{array}
\end{equation}
\begin{equation}
A\left ( q,k,v \right ) = Softmax(\frac{q\cdot k^{T} }{\sqrt{d_{k}} })\cdot v,
\end{equation}
where $X = \left \{ X_{1},X_{2},...,X_{n}  \right \}$ is the input sequence, $n$ is the length of the sequence, and $d_k$ is the dimension of the vectors, which also serves as the normalization factor to maintain the numerical stability of the attention scores. MHSA consists of multiple self-attentions, meaning the outputs from several self-attentions are integrated after mapping as:
\begin{equation}
\label{eq:multi head attention}
    \begin{array}{c}MHSA(X)=W^o\times (head_{1},head_{2},…,head_{i}),\\ \end{array}
\end{equation}
where $head_{i}$ is given as:
\begin{equation}
    head_{i}= A ( q_{i},k_{i},v_{i}).
\end{equation}

Following MHSA, there is a FFN layer, composed of fully connected networks. It performs a nonlinear mapping individually and identically on the output of MHSA at each position in the sequence. First, the features from the MHSA output undergo an upscaling projection to obtain high-dimensional feature representations. Then, these high-dimensional features are projected down to the original dimension size for seamless integration into the subsequent layer. The entire feed-forward network is as follows:
\begin{equation}
        FFN(X)= GeLU(XW_{1}+b_{1} ) W_{2}+b_{2},
\end{equation}
Among these, $W_{1} \in R^{D_{m}\times D_{f}}$, $b_{1}\in R^{D_{f}}$.$W_{2} \in R^{D_{f}\times D_{m}}$, and $b_{2} \in R^{D_{m}}$ are all trainable parameters. Typically, $D_{f}$ is larger than $D_{m}$, and generally $D_{f}$ is four times $D_{m}$. LN is applied after the residual connection, normalizing the output of the forward propagation to ensure that the output and input maintain a similar distribution. This helps to reduce the phenomenon of covariance shift and accelerate the convergence speed of network training.
\begin{equation}
        LN(X) = \frac{X-E(X)}{\sqrt{var[X]+\epsilon} }.
\end{equation}

In summary, the entire encoder network layer is defined as:
\begin{equation}
    \label{eq:encoder}
        \begin{split}
            & H^{'} = LN(MHSA(LN(X)+ LN(X))),\\
            & H = LN(FFN(H^{'})+ H^{'}).\\
        \end{split}
\end{equation}

Transformers can process entire sequences using a non-autoregressive approach. However, the self-attention mechanism within transformer networks is permutation invariant; for any given sequence, the attention scores remain the same regardless of the order of elements. However, the position within a sequence is often critical information. Therefore, in transformer networks, sinusoidal and cosine formulas of different frequencies are used across different dimensions to generate high-dimensional positional encodings that are added to the inputs. This method of incorporating positional information is called Positional Encoding, given as:
\begin{equation}
\label{eq:position enc}
    \left\{
    \begin{matrix}   
    PE(pos,2i)=sim(pos/10000^{2i/d_{model}}), \\   
    PE(pos,2i+1)=cos(pos/10000^{2i/d_{model}}),
    \end{matrix}
    \right. 
\end{equation}
where $i$ refers to the specific dimension within the data, $d_{model}$ represents the total dimensionality of the input data, and 10000 signifies that the model can accommodate sequences with a maximum length of 10000.

\subsection{Spatial-Spectral Network Module}
To enhance the representational power of features in EEG signal processing, we propose a novel spatial-spectral network. This network is designed to simultaneously extract significant spatial and frequency domain information from EEG signals at each time point, rather than processing these types of information separately. 

To realize this purpose, we optimize the multi-head attention component of the spatial-spectral network to handle both types of information simultaneously. Given an input signal $X \in R^{c \times f}$, where $c$ is the number of channels and $f$ is the number of frequency bands, the optimized multi-head self-attention mechanism maps the input signal $X$ to Query, Key, and Value vectors. During the multi-head self-attention computation, even-indexed heads process the channels as sequences for spatial feature extraction, while odd-indexed heads process the frequency bands as sequences for spectral feature extraction, as;
\begin{equation}
    \label{eq:opti self attention}
        \begin{array}{c}       
        \left\{\begin{matrix}   
            head_{i}(q_{i},k_{i},v_{i})=Softmax(\frac{q_{i}\cdot k_{i}^{T} }{\sqrt{d_{k}}})\cdot v_{i}, \\  
            head_{j}(q_{j},k_{j},v_{j})=Softmax(\frac{q_{j}^{T}\cdot k_{j} }{\sqrt{d_{k}}})\cdot v_{j}^{T}, \\  
            \end{matrix}\right.\end{array}
\end{equation}
Here, $i$ represents odd indices and $j$ represents even indices. The transformation of the Query, Key, and Value vectors in the transformer network is a linear mapping. Therefore, in the proposed spatial-spectral network, it is only necessary to transpose the Query, Key, and Value vectors of some heads to extract features from both spatial and spectral information, without the need to reconstruct the input sequences. Note that the inputs for spatial and spectral feature extraction are the sequences formed by the different channels of the input EEG signals at each time point and the extracted DE features at different frequency bands, respectively.

This proposed network simplifies the encoder process, reducing computational complexity while effectively capturing both spatial and spectral features. By transposing the Query, Key, and Value vectors for specific heads, the network could efficiently handle the different types of information. 

\subsection{Temporal Network Module}
The temporal network module receives input from the spatial-spectral network, which provides characterized spatial-spectral EEG information at each second. This information is represented as $L \in R^{t \times h}$, where $t$ denotes the time length and $h = c \times f$ represents the concatenation of channels and frequency bands. The resulting dimensionality of $h$, formed by combining channels and frequency bands, is typically high. To address this high dimensionality and create a more manageable feature space, we incorporate an embedding layer for dimensionality reduction before feeding the data into the next transformer. This embedding layer consists of a fully connected network that performs nonlinear mapping on the features at each second within the time series.

To facilitate sequence classification, we introduce a trainable parameter, CLS, which is randomly initialized and appended to the first position of the sequence. The CLS parameter plays a crucial role in extracting temporal information from the entire long-term EEG signal. Once the sequence, including the CLS parameter, is processed through the transformer's encoder, the final output CLS can effectively represent the entire long-term EEG signal. This approach leverages the power of the transformer architecture to capture intricate temporal patterns and dependencies within the EEG data, enhancing the model's ability to perform accurate and robust sequence classification.

In addition to the dimensionality reduction and sequence classification capabilities, the temporal network is designed to handle varying lengths of EEG signals by employing positional encoding. This encoding helps the model maintain the temporal order of the data, ensuring that the sequence information is preserved throughout the processing stages. 

\subsection{Transfer Learning Module}
One of the fundamental assumptions of deep learning is that training and testing data are drawn from the same distribution and are ideally independent and identically distributed (i.i.d.). However, EEG signals exhibit significant individual variability, making it challenging to satisfy this assumption in the task of emotion recognition from EEG signals. To address this issue, a transfer learning module is incorporated, which includes a feature extractor (Generator) and a domain discriminator (Discriminator). During training, the feature extractor and domain discriminator are trained in opposition to each other to learn domain-invariant features, effectively mitigating the impact of individual variability in EEG signals.

In the implementation, the subject data of the training set is defined as the source domain (Source), and the subject data of the test set is defined as the target domain (Target). The feature extractor is denoted as $G_{f}(\theta_{f})$, the domain discriminator as $G_{d}(\theta_{d})$, and the classifier as $G_{y}(\theta_{y})$. Here, $z$ represents the extracted features at the output of the temporal network, $y$ represents the label of the training set, and domain distribution alignment is achieved through adversarial loss:
\begin{equation}
    \begin{split}
         E(\theta_{f},\theta_{y},\theta_{d})=\sum_{z_{i}\in D_{s}}L_{y}(G_{y}(G_{f}(z_{i})),y_{i})-\\
        \lambda \sum_{z_{i}\in D_{s}\cup D_{t}}L_{d}(G_{d}(G_{f}(z_{i})),d_{i}), 
    \end{split}
\end{equation}
where $L_{y}$ is the classification loss, $L_{d}$ is the discriminator loss, and $d_{i}$ is the label for data domain. $d_{i}=0$ indicates that the data is from the source domain, while $d_{i}=1$ denotes that it is from the target domain. To facilitate implementation, a Gradient Reversal Layer (GRL) is incorporated into the network's backpropagation process. During forward propagation, GRL acts as an identity mapping:
\begin{equation}
        R_{\lambda }(z) = z.
\end{equation}

During backpropagation, GRL reverses the gradients:
\begin{equation}
        \frac{\mathrm{d} R_{\lambda}}{\mathrm{d} z}=-\lambda I, 
\end{equation}
where $\lambda$ is a dynamic balancing hyperparameter used to ensure the stability of the domain adversarial training process, given as:
\begin{equation}
        \lambda = \frac{2}{1+exp(-p)}-1.
\end{equation}

EEG signals from different subjects not only exhibit distributional differences but also individual temporal variations. To address this, our algorithm treats the spatial and temporal feature extraction networks as a unified feature extractor, rather than relying solely on the spatial feature extraction network. The loss function of the network is defined as follows:
\begin{equation}
    \label{eq:loss function}
        L_{overall} = L_{C}(c(f(X_{S})))+\lambda L_{adv}(d(f(X_{S},X_{T}))),
\end{equation}
where $L_{C}$ denotes the cross-entropy loss function for classification, $L_{adv}$ corresponds to the binary cross-entropy loss function for domain discrimination. $f(\cdot)$ is the feature extractor, $c(\cdot)$ is the classifier, and $d(\cdot)$ is the domain discriminator. $X_{S}$ and $X_{T}$ indicate the input data from the source and target domains, respectively. An overall algorithm of the proposed DuA transformer is illustrated in Algorithm \ref{alg:train}.

\begin{algorithm}
    \caption{The algorithm flow of DuA transformer.}
    \label{alg:train}
    \begin{algorithmic}[1]
        \Require
    \renewcommand{\algorithmicrequire}{\textbf{}}
    \Require - max iteration $\tau$, batch size $\xi$, length $t$, channel $c$, the size of feature $h$, source data $\{(X_s,Y_s)\}$, target data $\{X_t\}$.
    \Require - feature extractor $f(\cdot)$ contains spatial-spectral network $f_1(\cdot)$ and temporal network $f_2(\cdot)$, classifier $c(\cdot)$, discriminator $d(\cdot)$.
    \Ensure $f_1(\cdot)$, $f_2(\cdot)$, $c(\cdot)$, $d(\cdot)$.
    \State Random initialization of $f_1(\cdot)$,  $f_2(\cdot)$, $c(\cdot)$ and $d(\cdot)$;
    \For {$1$ to $\tau$} 
        \State Reshape $X = \{X_s,X_t\}$  $(\xi,t,c,h) \rightarrow (\xi \times t,c,h)$; 
        \State Generate second's feature $L=f_1(X)$;
        \State Reshape $L = \{L_s,L_t\}$  $(\xi \times t,c,h) \rightarrow (\xi,t,c \times h)$;
        \State Generate sequential feature $CLS=f_2(L)$;
        \State Calculate classification probability $q_c=c(CLS_s)$;
        \State Calculate discrimination probability  $q_d=d(CLS)$;
        \State Calculate overall loss $L_{overall}$;
        \State Gradient back-propagation;
        \State Update network parameters;
    \EndFor
    \end{algorithmic}
\end{algorithm}

\section{Experimental Results}
\label{sec:results}

\subsection{EEG Data and Experimental Protocol}
Given the limited availability of EEG emotional databases specifically designed for long-term EEG signals, we conducted a study to fill this gap by undertaking EEG emotional experiments under prolonged emotional stimulation. The study involved 50 subjects (25 males and 25 females, average age: 18.72 $\pm$ 1.23 years). A total of 12 long-term videos (6 positive and 6 negative) was selected as stimuli, each with an average duration of 11.11 $\pm$ 1.64 minutes. There was no content overlap among the selected videos. After watching each video, the subjects were asked to recall their emotions and rate their valence levels on a scale of [-10, -5, 0, 5, 10]. Here, -10 indicates extremely negative, 10 indicates extremely positive, and 0 refers to neutral. 

EEG signals were recorded in real-time using a Brain Products device with 63 channels configured according to the 10-20 system, as each subject watched the 12 long-term videos. To manage fatigue, the experiment was conducted in two sessions, with each session featuring videos that induced a single type of emotion (either positive or negative). These sessions were spaced at least one week apart. In total, the database comprises 600 long-term continuous EEG signal recordings. 

For the recorded EEG signals, we extracted DE features from each channel every second across the following ten frequency bands: Theta (4-6 Hz), Alpha1 (6-8 Hz), Alpha2 (8-10 Hz), Alpha3 (10-12 Hz), Beta1 (12-16 Hz), Beta2 (16-20 Hz), Beta3 (20-28 Hz), Gamma1 (28-34 Hz), Gamma2 (34-39 Hz), and Gamma3 (39-45 Hz). For each trial of the long-term continuous EEG signal recordings, the EEG signals were then transformed into a three-dimensional matrix by time, channel, and frequency band.

\begin{figure}[htbp]
    \centering
    \includegraphics[width=0.45\textwidth]{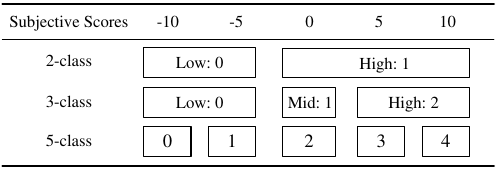}
    \caption{An illustration of different classification tasks.}
    \label{Fig:multi_task}
\end{figure}

For model validation, we adopt a trial-based leave-one-subject-out cross-validation protocol. Using our database of 50 subjects, we iteratively use the data from 49 subjects as the training set and the data from the remaining one subject as the test set. This process is repeated until each subject has served as the test set exactly once. The final validation results are obtained by calculating the average and standard deviation of these 50 test iterations. Additionally, leveraging the collected subjective scores ([-10, -5, 0, 5, 10]), we define three types of emotion classification tasks (2-class, 3-class, and 5-class), as illustrated in Fig. \ref{Fig:multi_task}. This approach ensures a robust assessment of the model's ability to generalize across different subjects and emotional states.

\subsection{Implementation Details}
In our experiment, we utilize the proposed DuA transformer, which integrates a spatial-spectral network and a temporal network as the feature extractor, and an MLP with a ReLU activation function as the domain discriminator. The parameters of the transformer network are randomly initialized from a uniform distribution. 

Specifically, the spatial-spectral network is configured as below. The input is $X \in R^{c\times f}$, where $c$ is the number of channels (63), $f$ is the number of frequency bands (10). The network uses 6 heads, with QKV vector mapping dimensions of 60 (each head having a QKV dimension of 10). The FFN layer is configured with a 10-40-10 mapping. The spatial-spectral network consists of one encoder layer. Subsequently, the channel and frequency band dimensions are concatenated and mapped from 630 to 128 before being fed into the temporal network. For the temporal network, it uses 3 heads, with QKV vector mapping dimensions of 384 (each head having a QKV dimension of 128). The FFN layer is configured with a 128-512-128 mapping. Similar as the spatial-spectral network, the temporal network also consists of one encoder layer.

During training, the number of epochs is set to 300. We use a random mini-batch gradient optimization algorithm to update the model parameters. The batch size for the training process is 12, the optimizer is Adam with a momentum parameter of 0.9, and the learning rate is $1\times10^{-3}$. The model is trained on an NVIDIA GeForce RTX 3090 GPU, using CUDA 10.0 drivers, and the training is conducted with the Pytorch API.

\begin{table}[h]
\caption{The mean accuracy (\%) and standard deviation (\%) of emotion recognition results under a comparison with the existing methods. '*' indicates that the experimental results are reproduced by ourselves. The \textbf{best results} are in bold, while the \underline{second-best results} are underlined.}
\centering
\label{tab:main_results}
\resizebox{\columnwidth}{!}{
\begin{tabular}{lccc}
\toprule
Methods  & \multicolumn{1}{c}{2-class}         & \multicolumn{1}{c}{3-class}         &\multicolumn{1}{c} {5-class}\\ \midrule
\multicolumn{4}{c}{\textit{Machine Learning}} \\ \midrule
SVM \cite{suykensLeastSquaresSupport1999}*      & 62.62±15.14 & 34.21±13.57 & 29.38±14.88 \\
KNN\cite{coomansAlternativeKnearestNeighbour1982}*      & 68.27±12.04 & 39.21±12.24 & \underline{32.70±16.11} \\
RF\cite{breimanRandomForests2001}*        & \underline{68.44±11.84} & \underline{39.88±11.15} & 28.24±15.77 \\
Adaboost\cite{hastieMulticlassAdaBoost2009}* & \underline{68.44±11.84} & 39.53±13.18 & 25.08±14.77 \\
TCA\cite{panDomainAdaptationTransfer2011}*      & 56.44±14.96 & 31.36±14.61 & 24.38±13.14 \\
SA\cite{fernandoUnsupervisedVisualDomain2013}*       & 65.44±15.80 & 39.06±11.70 & 23.74±11.22 \\
GFK\cite{gongGeodesicFlowKernel2012}*      & 55.27±12.06 & 32.74±11.96 & 24.08±13.33 \\
CORAL\cite{sunReturnFrustratinglyEasy2016}*    & 67.79±12.58 & 36.89±13.68 & 28.89±13.31 \\ \midrule
\multicolumn{4}{c}{\textit{Deap Learning}} \\ \midrule
DANet*      & 44.90±22.09 & 50.92±16.18 & 43.43±17.10 \\
MLP*        & 76.29±12.31 & 62.76±13.84 & 49.59±14.14 \\
DAN\cite{longLearningTransferableFeatures2015}*      & \underline{81.11±10.03} & 63.26±12.97 & 52.91±11.43 \\
DCORAL\cite{sunDeepCORALCorrelation2016}* & 79.44±10.36 & \underline{66.27±12.37} & 54.61±12.79 \\
DANN\cite{ganinDomainAdversarialTrainingNeural2016}*     & 80.27±11.72 & 65.59±12.95 & \underline{56.41±12.70} \\ 
\textbf{DuA (Ours)} & \textbf{85.27±08.56} & \textbf{76.77±08.87} & \textbf{64.43±13.10} \\ \midrule
Accuracy Gain  & +16.83/+04.16 & +36.89/+10.50 & +31.73/+08.02 \\ 
\bottomrule
\end{tabular}}
\end{table}

\subsection{Emotion Recognition Results}
We conduct an extensive evaluation of existing machine learning and deep learning methods on long-term continuous EEG signals using trial-based leave-one-subject-out cross-subject cross-validation. As shown in Table \ref{tab:main_results}, our proposed DuA transformer model demonstrates substantial performance improvements over traditional machine learning methods, achieving 16.83\% higher accuracy for 2-class classification, 36.89\% for 3-class classification, and 31.73\% for 5-class classification. When compared to advanced deep learning techniques, our DuA transformer also shows significant performance enhancements. Specifically, it outperforms the best transfer learning method by 4.16\% in 2-class classification, 10.50\% in 3-class classification, and 8.02\% in 5-class classification. These experimental results clearly demonstrate that the DuA transformer significantly surpasses both traditional machine learning and contemporary deep learning methods in trial-based long-term continuous EEG analysis. The superior performance of the DuA transformer can be attributed to its ability to effectively capture and utilize the complex spatial-spectral and temporal dependencies and patterns inherent in EEG signals. This capability enables it to provide more accurate and robust classification results across different classification tasks.

We further validate our proposed DuA transformer model on two benchmark databases: SEED \cite{zheng2015investigating} and SEED-IV \cite{zheng2018emotionmeter}. Unlike existing experiments on these databases, which predominantly employ \textbf{segment-based} leave-one-subject-out cross-validation, our study adopts a \textbf{trial-based} leave-one-subject-out cross-validation approach. This cross-validation shift enhances reliability, as emotional labels are assigned based on entire trials rather than segmented portions. This ensures a more accurate and holistic assessment of the model's performance in recognizing emotional states. The experimental results on the two benchmark databases are presented in Table \ref{tab:results on SEED} and Table \ref{tab:results on SEED-IV}, respectively.
\begin{table}
\caption{The mean accuracy (\%) and standard deviation (\%) of emotion recognition results on the SEED database.}
\centering
\label{tab:results on SEED}
\resizebox{\columnwidth}{!}{
\begin{tabular}{lccc}
\toprule
Methods  & \multicolumn{1}{c}{Session 1}         & \multicolumn{1}{c}{Session 2}        &\multicolumn{1}{c} {Session 3} \\ \midrule
DAN\cite{longLearningTransferableFeatures2015}* & 87.11±11.41 & 84.44±09.32 & 86.22±14.03 \\
DCORAL\cite{sunDeepCORALCorrelation2016}* & \underline{88.00±09.49} & \underline{84.88±10.12} & 85.77±10.76 \\
DANN\cite{ganinDomainAdversarialTrainingNeural2016}* & 87.11±07.25 & 81.33±12.06 & \underline{87.55±09.96} \\
\textbf{DuA (Ours)} & \textbf{91.55±05.99} & \textbf{87.55±09.38} & \textbf{88.44±08.64} \\ \midrule
Accuracy Gain  & +03.55 & +02.67 & +00.89 \\ 
\bottomrule
\end{tabular}}
\end{table}

\begin{table}
\caption{The mean accuracy (\%) and standard deviation (\%) of emotion recognition results on the SEED-IV database.}
\centering
\label{tab:results on SEED-IV}
\resizebox{\columnwidth}{!}{
\begin{tabular}{lccc}
\toprule
Methods  & \multicolumn{1}{c}{Session 1}         & \multicolumn{1}{c}{Session 2}        &\multicolumn{1}{c} {Session 3} \\ \midrule
DAN\cite{longLearningTransferableFeatures2015}* & 63.61±10.67 & 67.22±07.20 & 68.06±14.18 \\
DCORAL\cite{sunDeepCORALCorrelation2016}* & \underline{65.83±08.98} & \underline{70.28±08.05} & \underline{69.17±10.18} \\
DANN\cite{ganinDomainAdversarialTrainingNeural2016}* & 64.72±12.05 & 69.44±09.73 & 68.61±11.86 \\
\textbf{DuA (Ours)} & \textbf{72.22±08.28} & \textbf{78.33±07.42} & \textbf{72.50±12.38} \\ \midrule
Accuracy Gain  & +06.39 & +08.05 & +03.33 \\ 
\bottomrule
\end{tabular}}
\end{table}

In our analysis, we observe that the trial-based approach not only aligns more closely with real-world scenarios, where emotional experiences are continuous rather than discrete, but also addresses potential inconsistencies that may arise from segmenting data. By validating the model on these well-established databases using a more robust cross-validation technique, we demonstrate the DuA transformer's superior capability in handling complex emotional data. Additionally, our findings suggest that the trial-based cross-validation method provides a more comprehensive understanding of the model's effectiveness, particularly in dynamic and fluctuating emotional environments. This approach could significantly impact future research in emotion recognition, setting a new benchmark for evaluation practices. 

\section{Discussion Conclusion}
\subsection{Ablation Study}
We conduct ablation experiments to investigate the contributions of various components of our model to its overall performance. As shown in Table \ref{tab:ablation-spatspec}, we evaluate the role of the spatial-spectral network module under three conditions. (1) Complete Removal of the Spatial-Spectral Network (Temp-only). When the spatial-spectral network is entirely removed, there is a significant drop in accuracy: 8.83\% for 2-class classification, 7.66\% for 3-class classification, and 8.51\% for 5-class classification. This highlights the importance of the spatial-spectral network in achieving higher accuracy across different classification tasks. (2) Spatial Information Only (Spat-Temp). Retaining the spatial-spectral network but utilizing only spatial information results in a noticeable decline in performance across all classification tasks. This indicates that spatial information alone is insufficient for optimal model performance. (3) Spectral Information Only (Spec-Temp). Similarly, when we retain the spatial-spectral network but focus solely on spectral information, the results demonstrate that considering spatial information among EEG channels significantly benefits model performance. Balancing model performance and computational cost, our findings suggest that the proposed method strikes an effective balance. It not only reduces computational complexity but also preserves critical emotion representations, thereby maintaining robust performance across various classification tasks. This balanced approach ensures that the model remains both efficient and effective in recognizing and classifying emotions.
\begin{table}[h]
\centering
\caption{Ablation study of the spatial-spectral network module in our proposed DuA transformer. Temp-Only: complete removal of the spatial-spectral network; Spat-Temp: spatial information only; Spec-Temp: spectral information only.}
\color{black}
\label{tab:ablation-spatspec}
\resizebox{\columnwidth}{!}{%
\begin{tabular}{lcccc} 
\toprule
Methods & 2-class & 3-class & 5-class & Params\\
\midrule
Temp-Only &76.44±10.30 & 69.11±09.44 & 55.92±12.37 & 1.5K\\
Spat-Temp &82.28±10.39	&74.77±10.91	&60.77±12.02	& 122.52K\\
Spec-Temp &81.95±09.66	&73.10±11.25	&60.77±12.80	& 31.88K\\
\textbf{Full Model}     & \textbf{85.27±08.56} & \textbf{76.77±08.87} & \textbf{64.43±13.10} & \textbf{68.63K}\\
\bottomrule
\end{tabular}}
\end{table}

In the ablation experiments of the temporal network module, we conduct the experiments under two conditions to evaluate its significance (Table \ref{tab:ablation_temporal}). (1) Complete Removal of the Temporal Network (Spat-Spec w/o Temp). Eliminating the temporal network leads to a notable degradation in performance, with a 11.15\% decrease in 2-class classification, a 18.33\% decrease in 3-class classification, and a 20.02\% decrease in 5-class classification. This substantial drop underscores the critical role of temporal features in emotion recognition within long-term EEG signals. Long-term signals inherently contain more temporal information redundancy compared to short-term signals. The transformer network's self-attention mechanism allows it to focus on key time points, thereby extracting robust temporal features from these redundant long-term sequences. This finding demonstrates that neglecting temporal features significantly impairs the model's performance. (2) Retaining the Temporal Network but Removing Positional Encoding (Spat-Spec-Temp w/o PE). In this condition, we maintain the temporal network but remove the positional encoding from the transformer network. The results reveal that positional encoding is crucial for capturing the temporal dynamics of the signals. The absence of positional encoding results in a 14.50\% decrease in 2-class classification, a 11.51\% decrease in 3-class classification, and a 9.17\% decrease in 5-class classification. This decline highlights the importance of preserving the temporal order of signals in long-term EEG data, as it plays a vital role in the accurate classification of emotional states. Overall, these experiments illustrate the indispensable nature of temporal features and positional encoding in enhancing the performance of emotion recognition models utilizing long-term EEG signals. The findings emphasize that both the ability to capture temporal dependencies and the maintenance of signal order are essential for achieving robust and accurate emotion classification.

\begin{table}[h]
\centering
\caption{Ablation study of temporal network in our proposed DuA transformer. Spat-Spec w/o Temp: complete removal of the temporal network; Spat-Spec-Temp w/o PE: retaining the temporal network but removing positional encoding.}
\label{tab:ablation_temporal}
\resizebox{\columnwidth}{!}{
\begin{tabular}{lccc} 
\toprule
Methods & 2-class & 3-class & 5-class \\
\midrule
Spat-Spec w/o Temp &74.12±11.33	&58.44±11.57 &44.41±13.92 \\
Spat-Spec-Temp w/o PE &70.77±11.48 	&65.26±12.11	&55.26±12.65 \\
\textbf{Full Model}     & \textbf{85.27±08.56} & \textbf{76.77±08.87} & \textbf{64.43±13.10} \\
\bottomrule
\end{tabular}}
\end{table}

\subsection{Temporal Analysis Effect}
To further evaluate the effect of temporal analysis on long-term continuous EEG signals, we replace the temporal network module with an LSTM network. For a fair comparison, we only modify the temporal network in the proposed framework, maintaining the spatial feature extraction network unchanged to eliminate any experimental interference. As shown in Table \ref{tab:temp-LSTM}, our method significantly outperforms the LSTM method, improving recognition accuracy by 7.00\% in 2-class classification, 15.00\% in 3-class classification, and 12.84\% in 5-class classification.
\begin{table}[h]
\caption{Experimental comparison between LSTM and the temporal network module.}
\centering
\label{tab:temp-LSTM}
\begin{tabular}{llll}
 \toprule
Methods & \multicolumn{1}{c}{2-class}  & \multicolumn{1}{c}{3-class}         &\multicolumn{1}{c} {5-class} \\ \midrule
LSTM\cite{hochreiterLongShortTermMemory1997}*  & 78.27±10.72 & 61.77±11.03 & 51.59±11.39 \\
\textbf{DuA (Ours)}     & \textbf{85.27±08.56} & \textbf{76.77±08.87} & \textbf{64.43±13.10} \\ \midrule
Accuracy Gain  & \multicolumn{1}{c}{+07.00} & \multicolumn{1}{c}{+15.00} & \multicolumn{1}{c}{+12.84} \\ 
\bottomrule
\end{tabular}
\end{table}

The superior performance of our method can be attributed to the limitations of the LSTM network. LSTMs face issues with gradient accumulation due to their autoregressive calculation nature. As the length of the data sequence increases, the gradient accumulation becomes more pronounced, leading to the gradient vanishing problem. This issue hampers the LSTM's ability to effectively handle long-term signals. In contrast, the transformer network, which we employ in our method, does not rely on autoregressive calculations. Consequently, it avoids the gradient accumulation problem, making it more suitable for tasks involving long-term signals. The results further validate the advantages of using transformer networks over LSTMs for temporal analysis in long-term continuous EEG signal processing. Our findings highlight the importance of selecting appropriate temporal analysis methods to enhance efficiency and effectiveness in long-term continuous EEG emotion analysis.

\subsection{Hyperparameter Analysis}
\begin{figure*}[h]  
    \centering
    \includegraphics[width=0.75\textwidth]{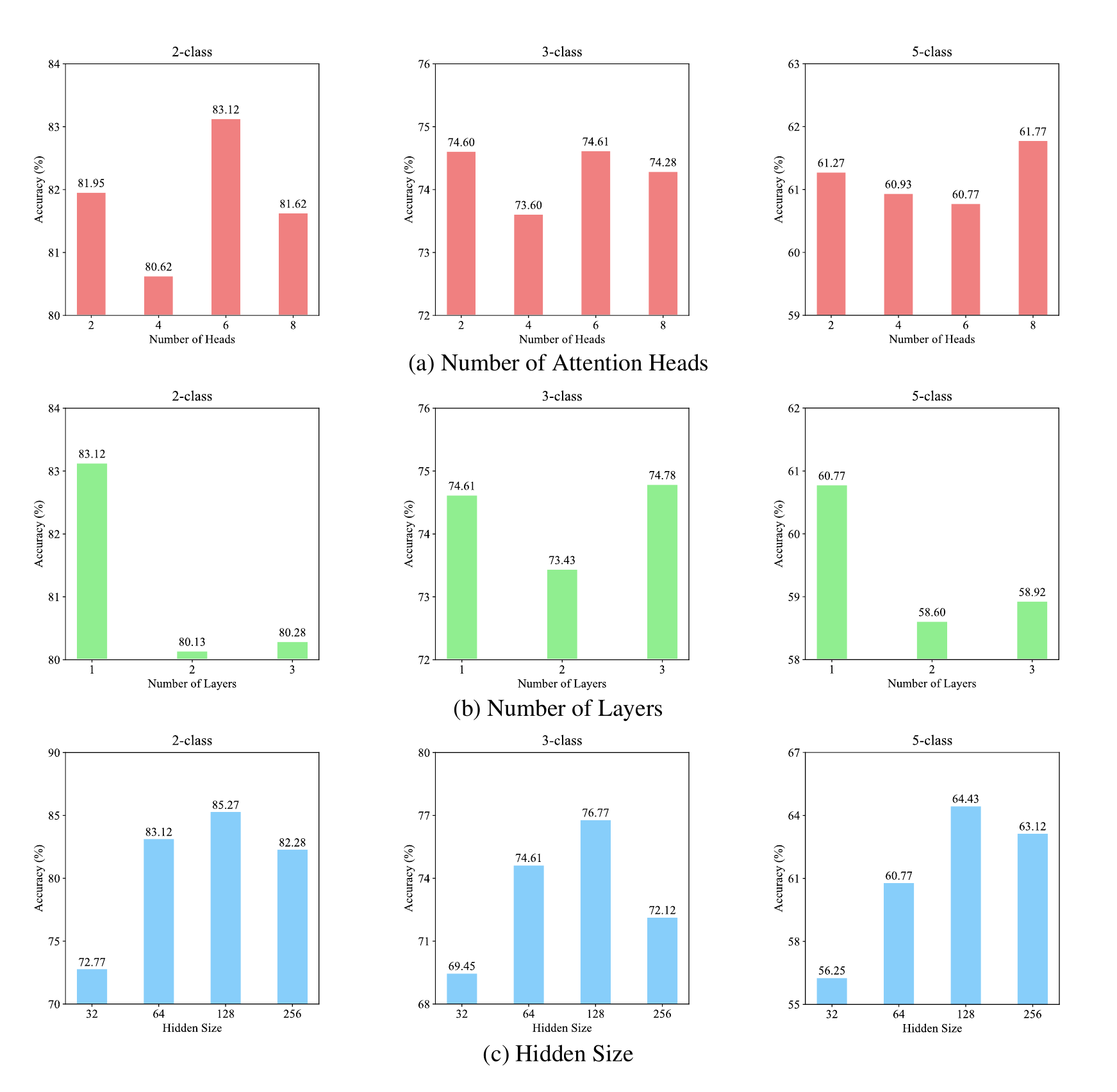}
    \caption{Hyperparameter analysis of (a) the number of attention heads in the spatial-spectral network, (b) the number of layers in the spatial-spectral network, and (c) the hidden size in the temporal network.}
    \label{Fig:hyperparam_analyse}
\end{figure*}

We conduct multiple sets of comparative experiments by adjusting the hyperparameters of our algorithmic framework. Given the constraints of computational power, we systematically analyze the model's sensitivity to various parameters. Specifically, we examine the number of heads in the spatial-spectral network (2, 4, 6, 8), the number of layers in the spatial-spectral network (1, 2, 3), and the hidden size in the temporal network (32, 64, 128, 256).

To ensure a robust analysis, we vary one set of hyperparameters while keeping the others constant. This method allows us to isolate the effects of each parameter on the model's performance. The experimental results, presented in Fig. \ref{Fig:hyperparam_analyse}, demonstrate that our model exhibits low sensitivity to changes in these hyperparameters. This stability indicates that the model is robust across a range of parameter values. Moreover, the results reveal that achieving optimal performance does not require extreme parameter values. Instead, moderate values are sufficient to attain the best outcomes, suggesting a balance between model complexity and computational efficiency. This finding is particularly important for practical applications where computational resources are limited.

\subsection{Conclusion}
In this paper, we present a novel Dual Attentive (DuA) transformer framework designed to address the challenges associated with processing long-term continuous EEG signals. Unlike existing segment-based processing pipelines, our approach treats a single trial as a comprehensive entity, analyzing dependencies across the entire trial by leveraging spatial-spectral and temporal information. The proposed DuA transformer outperforms various baseline methods, demonstrating the effectiveness of this optimized algorithm framework. By capturing the full scope of emotional experiences, our DuA transformer framework paves the way for the development of more accurate, reliable, and practical emotion recognition systems. Furthermore, the DuA transformer framework's ability to integrate and process comprehensive EEG data can lead to breakthroughs in understanding the neural underpinnings of emotions. This advancement holds significant potential for applications in mental health, human-computer interaction, and affective computing, offering improved tools for diagnosing and monitoring emotional states, enhancing user experience in interactive systems, and advancing research in affective sciences.

\section{Acknowledgments}
This work was supported in part by the National Natural Science Foundation of China under Grant 62276169, in part by the STI 2030 - Major Projects 2021ZD0200500, in part by the Medical-Engineering Interdisciplinary Research Foundation of Shenzhen University under Grant 2024YG008, in part by the Shenzhen University-Lingnan University Joint Research Programme, in part by the Shenzhen-Hong Kong Institute of Brain Science-Shenzhen Fundamental Research Institutions under Grant 2022SHIBS0003.


%




\ifCLASSOPTIONcaptionsoff
  \newpage
\fi

\bibliographystyle{IEEEtran}
\bibliography{references}



%


%

\begin{IEEEbiography}[{\includegraphics[width=1in,height=1.25in,clip,keepaspectratio]{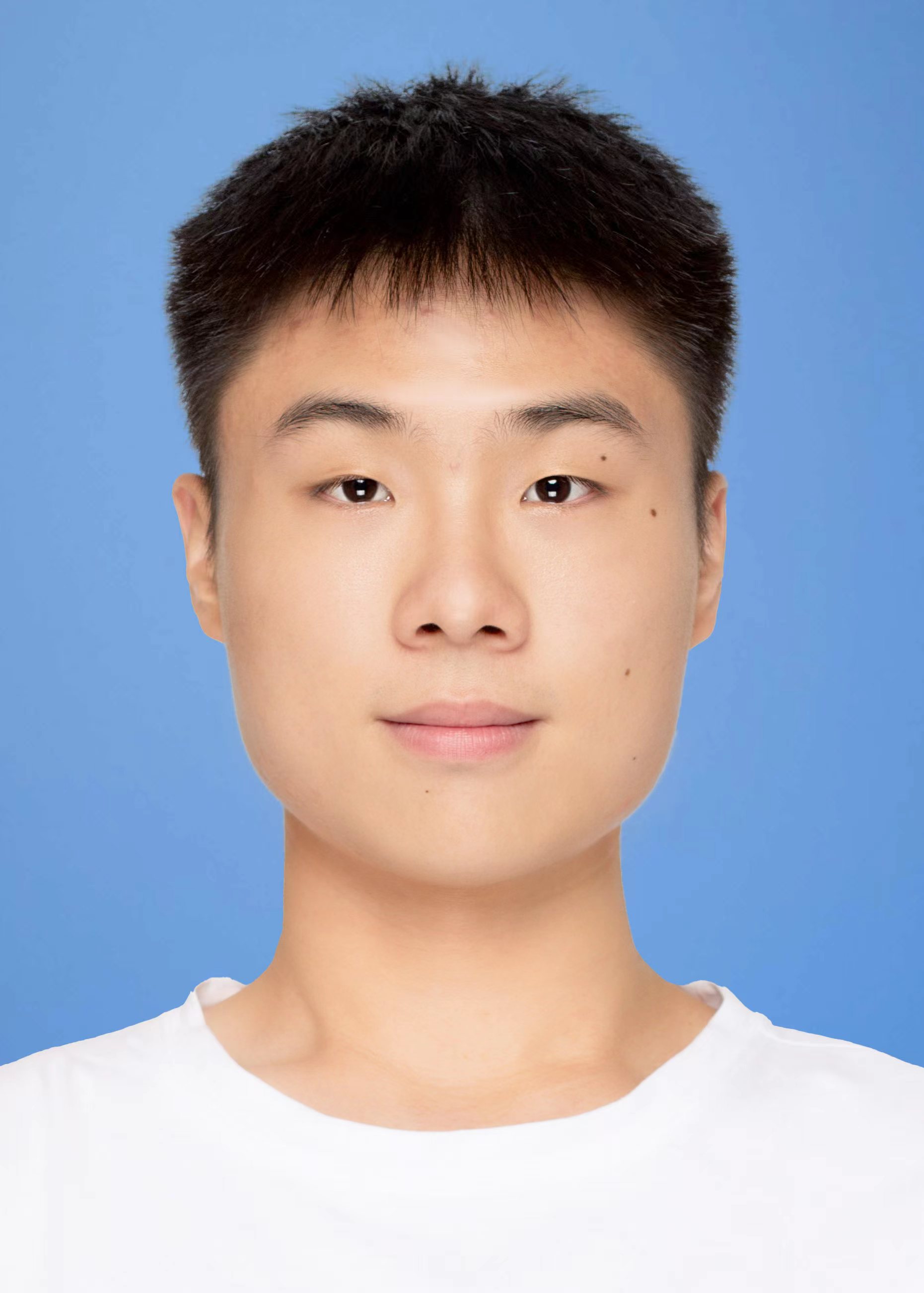}}]{Yue Pan}
received his B.E. degree from Shenzhen University in 2023. He is now a master in the Department of Biomedical Engineering, University of Electronic Science and Technology of China. His research interests include transfer learning and affective brain-computer interface.
\vspace{-10mm}
\end{IEEEbiography}

\begin{IEEEbiography}[{\includegraphics[width=1in,height=1.25in,clip,keepaspectratio]{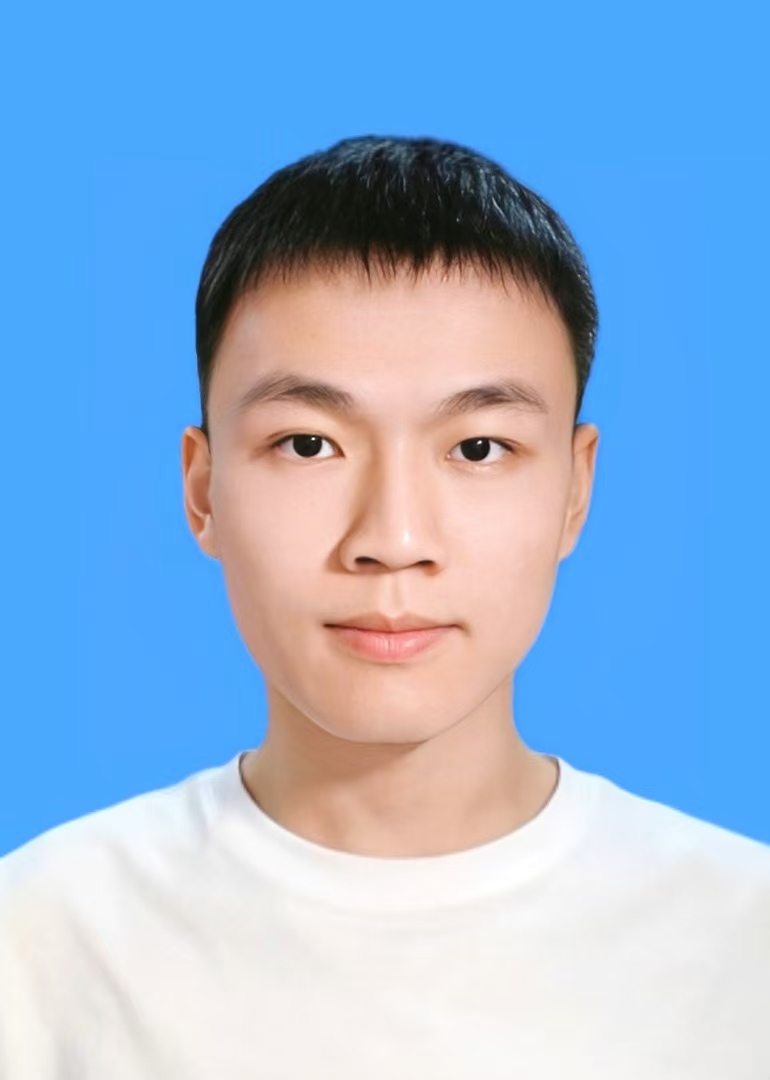}}]{Qile Liu}
received his B.E. degree from Guangdong University of Technology in 2022. He is currently pursuing the M.E. degree in the School of Biomedical Engineering, Medical School, Shenzhen University, China. His current research interests include brain-computer interface, affective computing, medical image analysis, and deep learning.
\vspace{-10mm}
\end{IEEEbiography}

\begin{IEEEbiography}[{\includegraphics[width=1in,height=1.25in,clip,keepaspectratio]{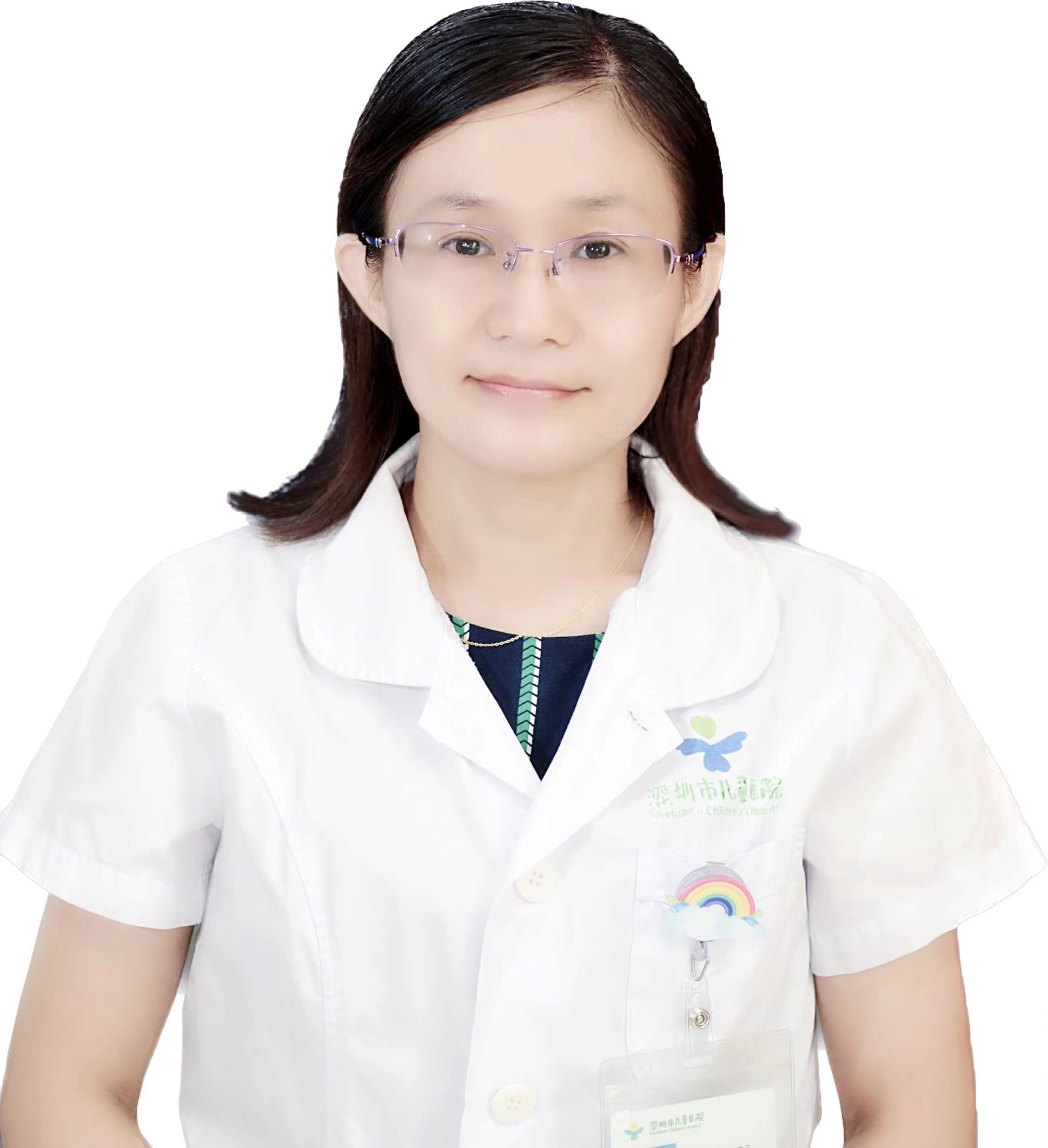}}]{Qing Liu}
received her M.D.degree from The Jiamusi university,Heilongjiang Province, China, in 2008. Now Studying at China Medical University.Since 2008, she has worked in Shenzhen Children's Hospital in China for 16years. She is currently the senior doctor of the Rehabilitation Department, Shenzhen Children’s Hospital,China. Her current research interests include brain injury and neurodevelopmental disorders in children.
\vspace{-10mm}
\end{IEEEbiography}

\begin{IEEEbiography}[{\includegraphics[width=1in,height=1.25in,clip,keepaspectratio]{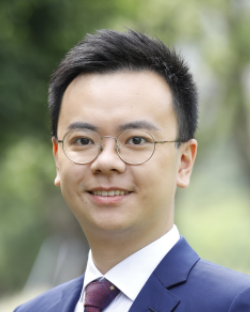}}]{Li Zhang}
received his Ph.D. degree from the Department of Electrical and Electronics, University of Hong Kong in 2017 and joined the School of Biomedical Engineering, Health Science Center, Shenzhen University in 2018 as an associate researcher. His current research interests mainly focus on biomedical signal processing, numerical optimization, and imaging genetics.
\vspace{-10mm}
\end{IEEEbiography}

\begin{IEEEbiography}[{\includegraphics[width=1in,height=1.25in,clip,keepaspectratio]{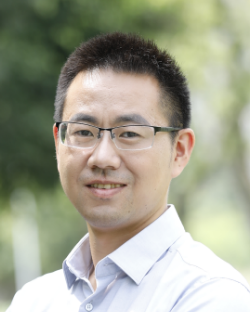}}]{Gan Huang}
received his Ph.D. degree from Shanghai Jiao Tong University, China (2013) and then received his postdoctoral training at Université Catholique de Louvain in Belgium. He is now an assistant professor at the School of Biomedical Engineering, Shenzhen University, China. He has published several high‐quality papers on journals including Neuroimage, Neural Network, and Neurocomputing. His current research interests focus on neural modulation, brain-computer interface, and neuroprosthesis.
\vspace{-10mm}
\end{IEEEbiography}

\begin{IEEEbiography}[{\includegraphics[width=1in,height=1.25in,clip,keepaspectratio]{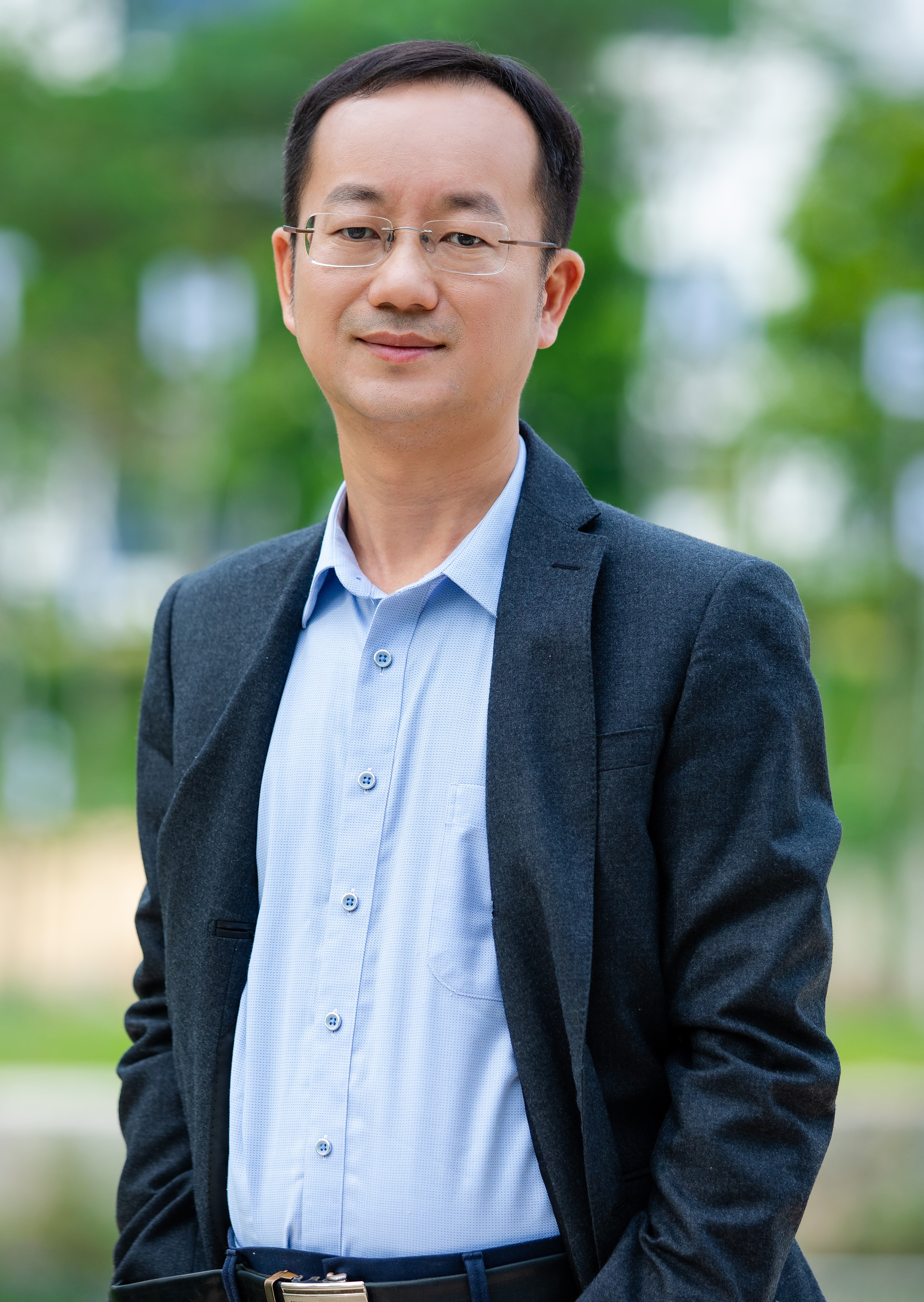}}]{Xin Chen} received his B.Sc. degree in electrical engineering and his Ph.D. degree in biomedical engineering from the University of Science and Technology of China in 1998 and 2003, respectively. He then joined the Department of Health Technology and Informatics at Hong Kong Polytechnic University as a research associate. In 2008, he joined the Department of Biomedical Engineering at Shenzhen University as an assistant professor and became an associate professor in 2011. He is currently the Dean and a professor at the School of Biomedical Engineering, Medical School, Shenzhen University, China. His main research interests include ultrasound imaging, ultrasound elasticity imaging and measurement, ultrasound image analysis, and biomedical instrumentation.
\vspace{-10mm}
\end{IEEEbiography}

\begin{IEEEbiography}[{\includegraphics[width=1in,height=1.25in,clip,keepaspectratio]{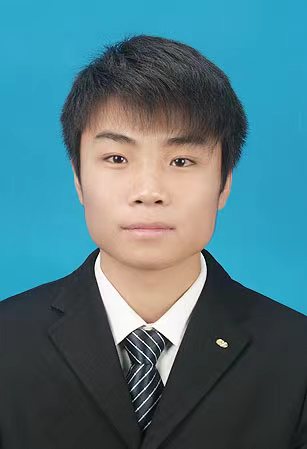}}]{Fali Li}
received his Ph.D degree from University of Electronic Science and Technology of China in 2020. He is now an associate researcher in the School of Life Science and Technology, Center for Information in Medicine, University of Electronic Science and Technology of China, China. His research interests include brain-computer interface, bio-informatics, and neural engineering.
\vspace{-10mm}
\end{IEEEbiography}

\begin{IEEEbiography}[{\includegraphics[width=1in,height=1.25in,clip,keepaspectratio]{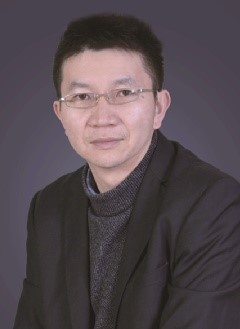}}]{Peng Xu}
received the Ph.D. degree in Biomedical Engineering from the School of Life Science and Technology, University of Electronic Science and Technology of China in 2006. He was a postdoctoral researcher at UCLA from 2007-02 to 2009-04. He is currently a full professor at the School of Life Science and Technology, University of Electronic Science and Technology of China, China. His research interests include brain–computer interface, brain inspired intelligence, machine learning, and brain network analysis, etc.
\vspace{-10mm}
\end{IEEEbiography}

\begin{IEEEbiography}[{\includegraphics[width=1in,height=1.25in,clip,keepaspectratio]{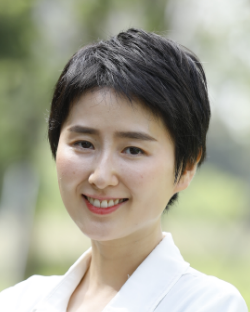}}]{Zhen Liang}
received her Ph.D. degree from The Hong Kong Polytechnic University, Hong Kong, in 2013. From 2012 to 2017, she was an algorithm development scientist at NeuroSky, Inc., Hong Kong. From 2018 to 2019, she was a specially‐appointed assistant professor of Graduate School of Informatics, Kyoto University, Japan. She is currently an associate professor in the School of Biomedical Engineering, Health Science Center, Shenzhen University, China. Her current research interests include brain encoding and decoding systems, affective computing, visual attention, and neural engineering.
\vspace{-10mm}
\end{IEEEbiography}












\end{document}